\documentclass[epj,nopacs,fleqn]{svjour}
\usepackage{epsfig,amsmath,amssymb,cite}
\usepackage{graphicx, subfigure}
\usepackage{slashed}
\usepackage{color}

\newcommand{\nn}{\nonumber}

\newcommand{\lam}{\lambda}
\newcommand{\sig}{\sigma}

\newcommand{\Mpl}{\overline{M}_{\rm Pl}}
\newcommand{\gld}{\tilde G}

\newcommand{\nored}{\tilde{\chi}}
\newcommand{\se}{\tilde e}

\newcommand{\psibar}{{\bar\psi}}

\newcommand{\Emiss}{\slashed{E}}
\def\del{\partial}

\newcommand{\idttheta}{{\int d^2 \theta \,}}
\newcommand{\idftheta}{{\int d^4 \theta \,}}
\newcommand{\hc}{{\text{h.c.}}}
\newcommand{\ubar}{\bar{u}}
\newcommand{\vbar}{\bar{v}}

\begin{document}
%%%%%%%%%%%% Begin Cover Page %%%%%%%%%%%%%%%%%%%%%%%%%%%%%%%%%%%%%%%%%%
\title{Monophoton signals in light gravitino production at $e^+ e^-$
colliders} 

\author{
 Kentarou Mawatari\inst{}\thanks{e-mail: kentarou.mawatari@vub.ac.be},
 Bettina Oexl\inst{}%\thanks{e-mail: bettina.oexl@vub.ac.be}
}

\institute{
 Theoretische Natuurkunde and IIHE/ELEM, Vrije Universiteit Brussel,\\
 and International Solvay Institutes,
 Pleinlaan 2, B-1050 Brussels, Belgium
 }

\abstract{
We revisit the monophoton plus missing energy signature at $e^+e^-$
colliders in supersymmetric (SUSY) models where the gravitino is very
light. There are two possible processes which provide the signal:
gravitino pair production and associated gravitino production with a
neutralino, leading the monophoton final state via an additional photon
radiation and via the neutralino decay, respectively. By using the
superspace formalism, we construct a model that allows us to study the
parameter space for the both processes. We show that the signal
cross section and the photon spectra provide information on the masses
of the SUSY particles as well as the SUSY breaking scale.  
}

%\date{\today}
\date{}

\titlerunning{Monophoton signals in light gravitino production at 
 $e^+e^-$ colliders} 
\authorrunning{K.~Mawatari, B.~Oexl}

\maketitle
%%%%%%%%%%%% End of Cover Page %%%%%%%%%%%%%%%%%%%%%%%%%%%%%%%%%%%%%%%%%

%%%%%%%%%%%%%% Begin Section 1 %%%%%%%%%%%%%%%%%%%%%%%%%%%%%%%%%%%%%%%%% 
\section{Introduction}\label{sec:intro}

Monophoton events with missing energy ($\gamma+\Emiss$) are one of
the promising search channels to find new physics at both lepton and
hadron colliders. So far no significant signal excess over the Standard
Model (SM) background has been observed at the 
LEP~\cite{Abbiendi:2000hh,Heister:2002ut,Achard:2003tx,Abdallah:2003np}
as well as at the 
Tevatron~\cite{Acosta:2002eq,Abazov:2008kp,Aaltonen:2008hh}
and the LHC~\cite{Chatrchyan:2012tea,Aad:2012fw}, constraining various
kinds of models, e.g. supersymmetry (SUSY) and extra dimensions.

The monophoton signal in the context of SUSY models has been searched
for models where the gravitino is the lightest SUSY particle (LSP) with
the very light mass 
$m_{3/2}\sim{\cal O}(10^{-14}-10^{-12}$~GeV) at the
LEP~\cite{Abbiendi:2000hh,Heister:2002ut,Achard:2003tx,Abdallah:2003np}
and the Tevatron~\cite{Acosta:2002eq}.% 
\footnote{A similar light-gravitino scenario has been studied in the
monojet plus missing energy signature ($j+\Emiss$) at the Tevatron~\cite{Affolder:2000ef} and the
LHC~\cite{ATLAS:2012zim}.}
In such scenarios there are two possible processes providing the signal: 
{\it gravitino pair production} ($\gld\gld$) and
{\it neutralino-gravitino associated production} ($\tilde\chi\gld$).
The former leads the monophoton final state via an additional photon
radiation, while the latter via the subsequent neutralino decay into a
photon and a LSP gravitino. 

The $\tilde\chi\gld$ associated production has been studied rather in
details~\cite{Fayet:1986zc,Dicus:1990vm,Lopez:1996gd,Lopez:1996ey,Baek:2002np,Mawatari:2011cu},
while the $\gld\gld(+\gamma)$ production has been investigated only in
models where all SUSY particles except for the gravitino are too heavy
to be produced on-shell~\cite{Nachtmann:1984xu,Brignole:1997sk,Brignole:1998me}. 

For the last few years simulation tools in the 
{\sc FeynRules}~\cite{Christensen:2008py,Duhr:2011se,Alloul:2013bka} and  
{\sc MadGraph}~\cite{Alwall:2007st,Alwall:2011uj} frameworks for
processes involving gravitinos/goldstinos have been intensively 
developed~\cite{Hagiwara:2010pi,Mawatari:2011jy,Christensen:2013aua},
making phenomenological studies 
easier~\cite{Mawatari:2011cu,Argurio:2011gu,deAquino:2012ru,Mawatari:2012ui,D'Hondt:2013ula,Ferretti:2013wya}.
It should be noted, however, that all the above recent studies
(except~\cite{Christensen:2013aua}) rely on the effective gravitino
Lagrangian that contains only interactions with a single gravitino. To
study the $\gld\gld$ production, we need a consistent implementation of
all the relevant interactions including vertices involving two
gravitinos as well as sgoldstinos, which are the superpartners of
goldstinos and play an important role for the
unitarity~\cite{Bhattacharya:1988ey,Bhattacharya:1988zp}. We also note
that the process contains a four-fermion interaction involving two
Majorana particles, which is not supported in the default
{\sc MadGraph}, and therefore special implementations are required.

In this article, we consider a scenario where the gravitino is the LSP
and the lightest neutralino is the next-to-lightest SUSY particle (NLSP)
and promptly decays into a photon and a gravitino. We revisit the
monophoton plus missing energy signature for future $e^+e^-$ colliders
\begin{align}
 e^+e^-\to\gamma\gld\gld\to\gamma+\Emiss,
\end{align}
where, as mentioned, the $\gld\gld$ and $\tilde\chi\gld$ productions can
be the dominant subprocesses. In order to study the whole parameter
space for the both processes, including all the relevant SUSY particles
as well as sgoldstinos, we construct a simple SUSY QED model with a
goldstino multiplet in the gravitino-goldstino equivalence limit by
using the superspace formalism. We investigate the $e^+e^-\to\gld\gld$
process in detail to see how the cross section deviates from that in
models where all SUSY particles except for the gravitino are assumed to
be heavy and integrated out. We generate the signal samples as well as
the SM background, and analyze the signal cross sections and the photon 
spectra to extract information on the masses of the neutralino and
selectrons as well as the gravitino mass, which is related to the SUSY
breaking scale. 

We note in passing that, although our study in this article focuses on
lepton colliders, all the results are applicable for $\gamma+\Emiss$ as
well as jet$+\Emiss$ signals at hadron colliders and the detailed study
will be reported elsewhere. 

The paper is organized as follows: 
In Sect.~\ref{sec:model} we construct a SUSY QED model including
interactions with (s)goldstinos in the superspace formalism.  
In Sect.~\ref{sec:gldpair}, we explore the parameter space in the
$e^+e^-\to\gld\gld$ process, and briefly review the
$e^+e^-\to\tilde\chi\gld$ process. 
In Sect.~\ref{sec:signal}, we simulate the $e^+e^-\to\gamma\gld\gld$
process as well as the SM background, and show that the signal cross
sections and the photon spectra provide information on the masses of the
neutralino and selectrons as well as the gravitino mass.
Sect.~\ref{sec:summary} is devoted to our summary. 
In Appendix~\ref{sec:lag} we give the relevant Lagrangian in terms of
the component fields.
In Appendix~\ref{sec:aa}, to validate our model implementation of
sgoldstinos, we briefly discuss the $\gamma\gamma\to\gld\gld$ process.

%%%%%%%%%%%%%% Begin Section 2 %%%%%%%%%%%%%%%%%%%%%%%%%%%%%%%%%%%%%%%%% 
\section{SUSY QED with a goldstino superfield}\label{sec:model}

In phenomenologically viable SUSY models, the SUSY breaking is usually
assumed to happen in a so-called hidden sector and then being
transmitted to the visible sector (i.e. the SM particles and their
superpartners) through some mediation mechanism. As a result, one
obtains effective couplings of the fields in the visible sector to the
goldstino multiplet. To illustrate the interactions among the physical
degrees of freedom of the goldstino multiplet and the fields in the
visible sector, we discuss an $R$-parity conserving $N=1$ global
supersymmetric model with the $U(1)_{\text{em}}$ gauge group in the
superspace formalism. The model comprises one vector superfield
$V=(A^{\mu},\lambda,D_V)$, describing a photon $A^{\mu}$ and a photino
$\lambda$, and two chiral superfields $\Phi_{L}=(\se_{L},e_{L}, F_{L})$
and $\Phi_{R}=(\se^*_{R},e_{R}^{c},F_{R})$, containing the left- and 
right-handed electrons $e_{L/R}$ and selectrons $\se_{L/R}$. In
addition, we introduce a chiral superfield in the hidden sector
$X=(\phi,\gld,F_X)$, containing a sgoldstino $\phi$ and a goldstino
$\gld$. $D_V$, $F_{L/R}$ and $F_X$ are auxiliary fields.

The Lagrangian of the visible sector is 
\begin{align}
 \mathcal{L}_{\rm vis}=
 &\sum_{i=L,R}\idftheta\,\Phi^{\dagger}_ie^{2g_eQ_iV}\Phi_i \nn\\
 &+\frac{1}{4}\Big(\idttheta\,W^{\alpha} W_{\alpha} +\hc\Big),
\label{L_vis}
\end{align}
where $g_e=\sqrt{4\pi\alpha}$ and $Q_i$ is the electric charge of
$\Phi_i$, i.e. $Q_{R/L}=\pm 1$.%  
\footnote{The covariant derivative is defined as
$D_{\mu}=\del_{\mu}+ig_eQA_{\mu}$.}
$W_{\alpha}=-\frac{1}{4}\bar{D}\cdot\bar{D}D_{\alpha}V$ denotes the SUSY
$U(1)_{\text{em}}$ field strength tensor with $D$ being the
superderivative. $\mathcal{L}_{\rm vis}$ contains the kinetic terms as
well as the gauge interactions. 

The Lagrangian of the goldstino superfield is given by
\begin{align}
 \mathcal{L}_{X}=
 &\idftheta\,X^{\dagger}X-\Big(F\idttheta\,X+\hc\Big) \nn\\
 &-\frac{c_X}{4}\idftheta\,(X^{\dagger}X)^2.
\label{L_hid}
\end{align}
The first term gives the kinetic term of the (s)goldstino, while
the second term is a source of SUSY breaking and 
$F\equiv\langle F_X\rangle$ is a vacuum expectation value (VEV) of
$F_X$.% 
\footnote{Note that we follow the {\sc FeynRules} convention for chiral 
superfields 
$\Phi(y,\theta)=\phi(y)+\sqrt{2}\,\theta\cdot\psi(y)-\theta\cdot\theta\,F(y)$~\cite{Alloul:2013bka}, 
which fixes the sign of the Lagrangian so as to give a positive
contribution to the scalar potential.} 
The last term is non-renormalizable and provides interactions between
the goldstino multiplet. This term also gives the sgoldstino mass term
when replacing the auxiliary fields $F_X$ by the VEV, and hence we
assign $c_X=m^2_{\phi}/F^2$. 

The interactions among the (s)goldstinos and the fields in the visible
sector as well as the soft mass terms for the selectrons and the photino
are given by the effective Lagrangian 
\begin{align}
 \mathcal{L}_{\text{int}}=&-\sum_{i=L,R}c_{\Phi_i}\idftheta\, 
  X^{\dagger}X\Phi_i^{\dagger}\Phi_i\nn\\
 &-\Big(\frac{c_V}{4}\idttheta\, X W^{\alpha} W_{\alpha} +\hc\Big),
\label{L_int}
\end{align}
where we identify $c_{\Phi_i}=m^2_{\se_i}/F^2$ and $c_V=2m_{\lambda}/F$. 

We note that our model is minimal, yet enough to investigate the
$\gamma+\Emiss$ signal at $e^+e^-$ colliders.
We also note that our Lagrangian is model independent.
However, studies of non-linear SUSY revealed that additional
model dependent terms for four-point effective interactions involving
two goldstinos and two matter fermions are
allowed~\cite{Luty:1998np,Brignole:1997pe,Clark:1997aa}. One possible 
source for such terms is $D$-type SUSY breaking~\cite{Brignole:2003cm},
which does not occur in our model. 

Before turning to collider phenomenology, we briefly refer to the
goldstino equivalence theorem. When the global SUSY is promoted to the
local one, the goldstino is absorbed by the gravitino via the
super-Higgs mechanism. 
In the high-energy limit, $\sqrt{s}\gg m_{3/2}$, which is always
fulfilled for very light gravitinos at colliders, the interactions of
the helicity 1/2 components are dominant, and can be well described by
the goldstino interactions due to the graviton-goldstino equivalence
theorem~\cite{Casalbuoni:1988kv,Casalbuoni:1988qd}. 
We also note that, as a
consequence of the super-Higgs mechanism, the gravitino mass is related
to the scale of the SUSY breaking and the Planck scale, in a flat
space-time, as~\cite{Volkov:1973jd,Deser:1977uq}
\begin{align}
 m_{3/2}=\frac{F}{\sqrt{3}\,\Mpl},
\label{grav_mass}
\end{align}
where $\Mpl\equiv M_{\rm Pl}/\sqrt{8\pi}\approx 2.4\times10^{18}$~GeV is
the reduced Planck mass. Therefore, low-scale SUSY breaking scenarios
provide a gravitino LSP. In the following, we simply call the goldstino
the gravitino and also call the photino the (lightest) neutralino
$\tilde\chi$. We note that by construction we ignore other neutralino
mixing scenarios. 
Since the zino and higgsino mixing gives rise to the
$Z$ and $H$ decay modes of the neutralino~\cite{Ambrosanio:1996jn}, the
overall $\gamma+\Emiss$ rate decreases, but the property of the signal
does not change. 
The extension of our model to the SM gauge
group is straightforward to study the general minimal supersymmetric SM
(MSSM); see e.g.~\cite{Antoniadis:2010hs}. 

For completeness, we show the relevant interaction Lagrangians
of~\eqref{L_vis}, 
\eqref{L_hid} and \eqref{L_int} in terms of the component fields in
Appendix~\ref{sec:lag}. We have implemented the above Lagrangian by
using the superspace module into
{\sc FeynRules~2}~\cite{Alloul:2013bka}, which provides the Feynman
rules in terms of the physical component fields and the {\sc UFO} model
file~\cite{Degrande:2011ua,deAquino:2011ub} for matrix-element
generators such as {\sc MadGraph 5}~\cite{Alwall:2011uj}.

%%%%%%%%%%%%%% Begin Section 3 %%%%%%%%%%%%%%%%%%%%%%%%%%%%%%%%%%%%%%%%% 
\section{Light gravitino production at $e^+e^-$ colliders}
\label{sec:gldpair}

Based on the model we constructed in the previous section, we
investigate direct LSP gravitino production processes that lead to
$\gamma+\Emiss$ at future $e^+e^-$ colliders. We consider the neutralino
to be the NLSP and to promptly decay into a photon and a gravitino. The
missing energy will be carried away by two gravitinos due to the
$R$-parity conservation. Two distinct processes give rise to the signal: 
\textit{gravitino pair production} ($\gld\gld$) and
\textit{neutralino-gravitino associated production} ($\nored\gld$),
leading the monophoton final state via an additional photon radiation
and via the subsequent neutralino decay, respectively. Their relative
importance varies with the gravitino and neutralino masses as well as
with kinematical cuts. In the following, a detailed discussion of the
$\gld\gld$ production is presented, followed by a short review of the
$\nored\gld$ production. According to the cross sections, we fix the
benchmark points for our simulation in the next section. We also comment
on the validation of our model implementation in the last part of this
section.

\subsection{Gravitino pair production}\label{sec:grav_pair_prod}

\begin{figure}
\center
 \includegraphics[width=0.32\columnwidth]{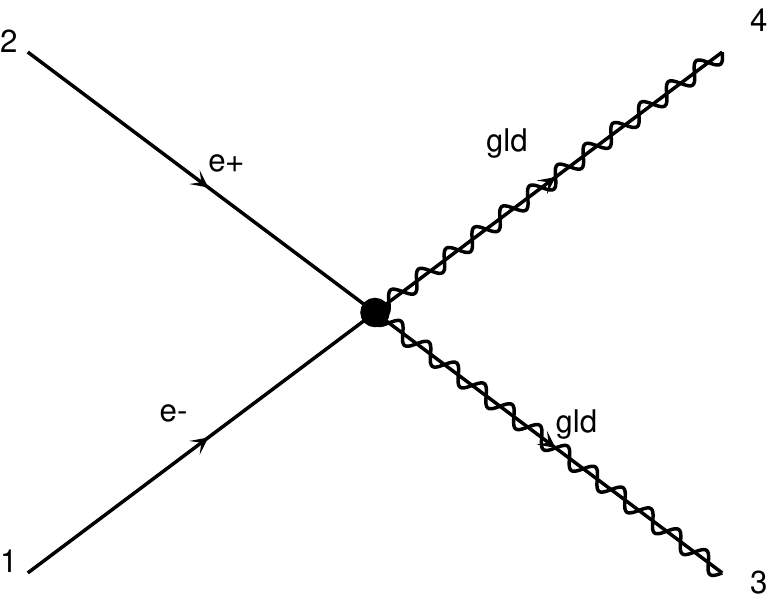}
 \includegraphics[width=0.32\columnwidth]{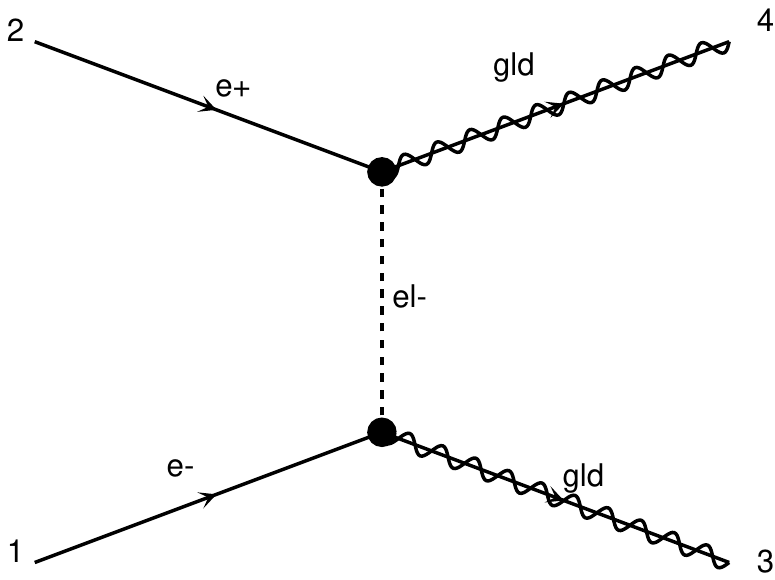}
 \includegraphics[width=0.32\columnwidth]{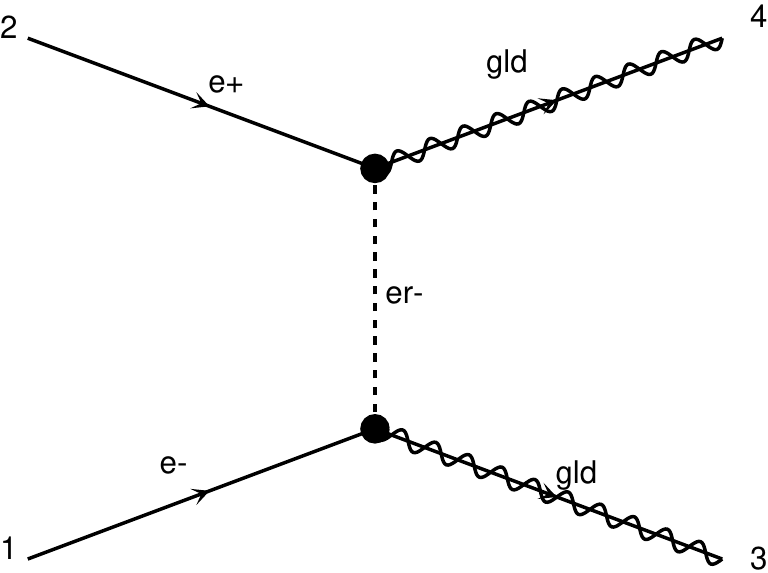}
 \caption{Samples of Feynman diagrams for gravitino pair
 production in $e^+e^-$ collisions, generated by (modified) 
 {\sc MadGraph 5}~\cite{Alwall:2011uj}. {\tt gld}, {\tt el}, and
 {\tt er} denote a gravitino, a left-handed selectron, and a
 right-handed selectron, respectively.}
\label{fig:diagram}
\end{figure}

Gravitino pair production gives rise to the monophoton plus missing
energy signature when an additional photon is
emitted~\cite{Nachtmann:1984xu,Brignole:1997sk}. Here we present the
helicity amplitudes explicitly for the two-to-two process 
\begin{align}
  e^-\Big(p_1,\frac{\lambda_1}{2}\Big)
 +e^+\Big(p_2,\frac{\lambda_2}{2}\Big)
 \to
  \gld\Big(p_3,\frac{\lambda_3}{2}\Big) 
 +\gld\Big(p_4,\frac{\lambda_4}{2}\Big),
\label{epluseminus_gldgld}
\end{align}
where the four momenta ($p_i$) and helicities ($\lambda_i=\pm1$) are
defined in the center-of-mass (CM) frame of the $e^+e^-$ collision. 
In the massless limit of $e^{\pm}$, one can find that all amplitudes are
zero when both the electron and the positron have the same helicity, and
hence we fix $\lambda_2=-\lambda_1$. The same helicity relation holds
for the massless gravitinos in the final state, leading to
$\lambda_4=-\lambda_3$. Since we will assume gravitinos with mass
$m_{3/2}\sim\mathcal{O}(10^{-13}{\rm ~GeV})$, we neglect the gravitino 
mass in the phase space but keep it in the couplings. In addition, for
the $\lambda_1=+1$ ($\lambda_1=-1$), only right-handed (left-handed)
selectrons can contribute to the total amplitudes. Therefore, the
helicity amplitudes for the above process can be expressed as the sum of
the four-point contact amplitude and the $t,u$-channel selectron
exchange amplitudes (see also Fig.~\ref{fig:diagram}): 
\begin{align}
  \mathcal{M}^{}_{\lambda_1,\lambda_3}
 =\mathcal{M}^c_{\lambda_1,\lambda_3}
 +\mathcal{M}^t_{\lambda_1,\lambda_3}
 +\mathcal{M}^u_{\lambda_1,\lambda_3}.
\label{amp_ee}
\end{align}
Using the straightforward Feynman rules for Majorana fermions given 
in~\cite{Denner:1992vza}, the above amplitudes are written, based on the 
effective gravitino Lagrangian in Appendix~\ref{sec:lag}, as 
\begin{align}
  i\mathcal{M}^c_{\lambda_1,\lambda_3}
 &=-\frac{im^2_{\se_{\lambda_1}}}{F^2}
   \big( \hat{\mathcal{M}}^t_{\lambda_1,\lambda_3}
        -\hat{\mathcal{M}}^u_{\lambda_1,\lambda_3}\big), \\ 
  i\mathcal{M}^t_{\lambda_1,\lambda_3}
 &=-\frac{im^4_{\se_{\lambda_1}}}{F^2(t-m^2_{\se_{\lambda_1}})}\,
   \hat{\mathcal{M}}^t_{\lambda_1,\lambda_3}, \\
  i\mathcal{M}^u_{\lambda_1,\lambda_3}
 &=\frac{im^4_{\se_{\lambda_1}}}{F^2(u-m^2_{\se_{\lambda_1}})}\,
   \hat{\mathcal{M}}^u_{\lambda_1,\lambda_3},
\end{align}
where $m_{\se_{\pm}}$ denotes the right/left-handed selectron mass for
notational convenience. The reduced helicity amplitudes are 
\begin{align}
  \hat{\mathcal{M}}^t_{\lambda_1,\lambda_3}
 &=\ubar(p_3,\lambda_3)P_{\lambda_1}u(p_1,\lambda_1) \nn\\
 &\quad\times \vbar(p_2,-\lambda_1)P_{-\lambda_1}v(p_4,-\lambda_3),\nn\\
  \hat{\mathcal{M}}^u_{\lambda_1,\lambda_3}
 &=\ubar(p_4,-\lambda_3)P_{\lambda_1}u(p_1,\lambda_1) \nn\\
 &\quad\times\vbar(p_2,-\lambda_1)P_{-\lambda_1}v(p_3,\lambda_3),
\end{align}
where $P_{\pm}=\frac{1}{2}(1\pm\gamma^5)$ is the chiral projection
operator.

\begin{table*}
\center
 \caption{The helicity amplitudes $\mathcal{M}_{\lambda_1,\lambda_3}$
 defined in~\eqref{amp_ee} for 
 $e^-_{\lambda_1}e^+_{-\lambda_1}\to\gld^{}_{\lambda_3}\gld^{}_{-\lambda_3}$.}
\begin{tabular}{c|cccccc}
\hline
 $\lambda_1\lambda_3$&&& $\mathcal{M}^c$&$\mathcal{M}^t$ & $\mathcal{M}^u$&\\
\hline
$\pm\,\mp$&$-\dfrac{s\,m^2_{\se_{\lambda_1}}}{2F^2}\, (1-\cos{\theta})$&$\Big[$&$1$&$+\dfrac{m^2_{\se_{\lambda_1}}}{t-m^2_{\se_{\lambda_1}}}$&&$\Big]$
 \\ 
$\pm\,\pm$&$-\dfrac{s\,m^2_{\se_{\lambda_1}}}{2F^2}\,(1+\cos{\theta})$&$\Big[$&$1$&&$+\dfrac{m^2_{\se_{\lambda_1}}}{u-m^2_{\se_{\lambda_1}}}$&$\Big]$ \\
\hline
\end{tabular}
\label{tb:helamp_ee}
\end{table*}

\begin{figure}
\center
 \includegraphics[width=0.88\columnwidth]{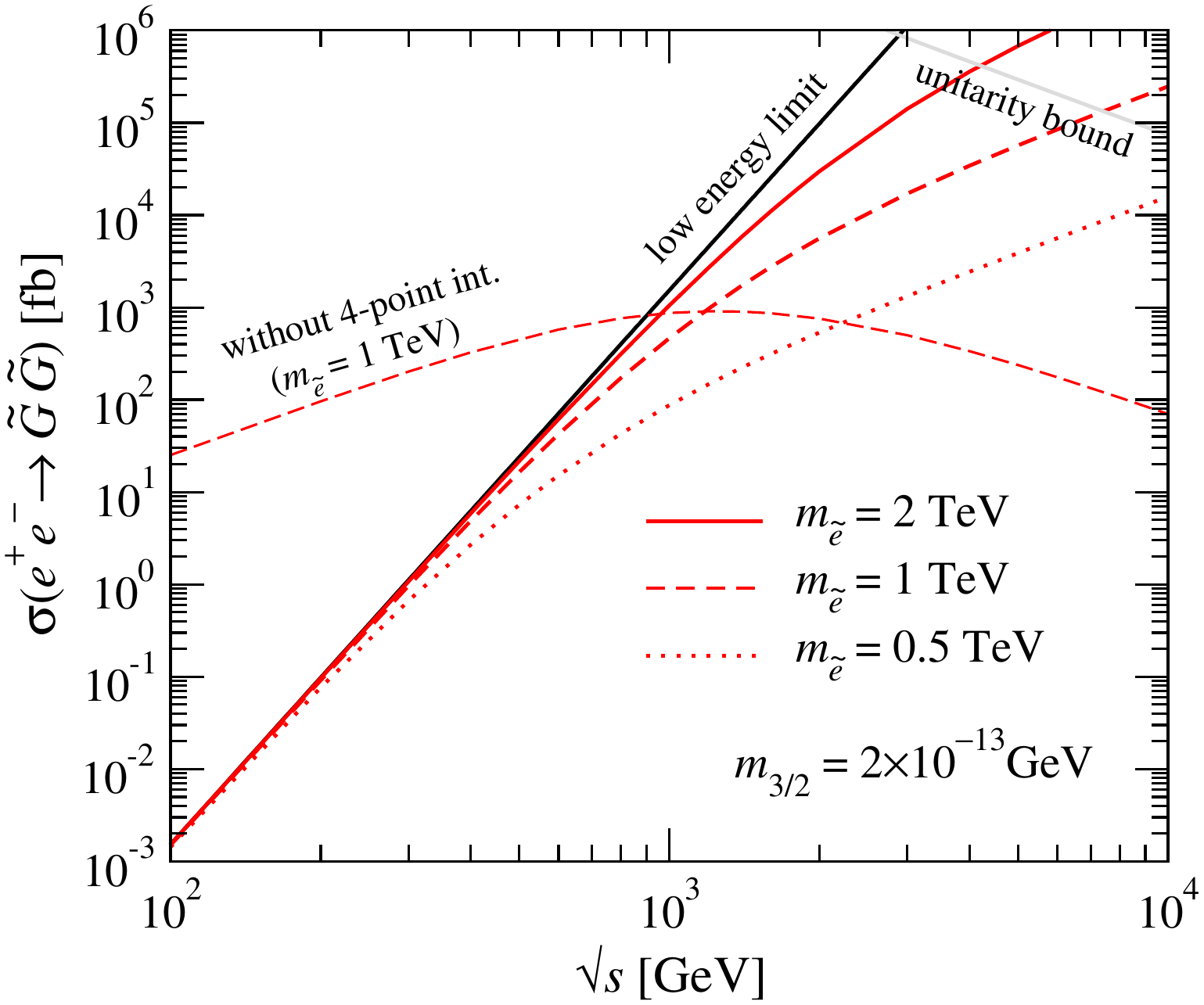}
 \caption{Total cross sections of
 $e^+e^-\to\gld\gld$ as a function of the collision energy for different
 selectron masses $m_{\se_{\pm}}=0.5,1,2$~TeV with
 $m_{3/2}=2\times10^{-13}$~GeV.
 The cross section in the low-energy limit is presented by a black solid
 line. The contribution without the four-point interaction for 
 $m_{\se_{\pm}}=1$~TeV is also shown as a reference.}
\label{fig:xsec_rs} 
\end{figure}

With the four momenta defined as
\begin{align}
 p_1^{\mu}=&\frac{\sqrt{s}}{2}(1,0,0,1),\nn\\
 p_2^{\mu}=&\frac{\sqrt{s}}{2}(1,0,0,-1),\nn\\
 p_3^{\mu}=&\frac{\sqrt{s}}{2}(1,\sin{\theta},0,\cos{\theta}),\nn\\
 p_4^{\mu}=&\frac{\sqrt{s}}{2}(1,-\sin{\theta},0,-\cos{\theta}),
\end{align}
we present the helicity amplitudes in Table~\ref{tb:helamp_ee}. The
total cross section is given by 
\begin{align}
 \sigma
 &=\frac{1}{192\pi F^4}
 \sum_{\lambda=\pm}\frac{m_{\se_{\lambda}}^4}{s^2}
  \bigg[ s^3 
       -3m_{\se_{\lambda}}^2 s^2 
       +9m_{\se_{\lambda}}^4 s \nn\\
 &\hspace*{9mm}+3m_{\se_\lambda}^{6}
  \Big( 1-\frac{m^2_{\se_{\lambda}}}{s+m^2_{\se_{\lambda}}}
       +4\log{\frac{m_{\se_{\lambda}}^2}{s+m_{\se_{\lambda}}^2}}\Big)
 \bigg].
\label{xsec_gldgld}
\end{align}
Figure~\ref{fig:xsec_rs} shows the total cross sections as a function of
the CM energy $\sqrt{s}$ for three different selectron masses 
$m_{\se_{\pm}}=0.5$, 1 and 2~TeV. The gravitino mass is fixed at 
$m_{3/2}=2\times10^{-13}$~GeV, which corresponds by~\eqref{grav_mass}
to the SUSY breaking scale $\sqrt{F}\approx 918$~GeV. We stress that the
cross section is extremely sensitive to the gravitino mass since it
scales inversely proportionally to the gravitino mass to the fourth,
\begin{align}
 \sigma(\gld\gld)\propto 1/m_{3/2}^4.
\label{sig_gldgld}
\end{align}
We also note that the cross section tends to be larger for the heavier
selectrons since the couplings are proportional to $m_{\se}^2$.

In the low-energy limit, $\sqrt{s}\ll m_{\se_{\pm}}$, as one can easily
see from the explicit amplitudes in Table~\ref{tb:helamp_ee}, a strong
cancellation happens between $\mathcal{M}^c$ and $\mathcal{M}^{t,u}$,
leading to a cross section scaling
as~\cite{Brignole:1997pe,Brignole:1997sk} 
\begin{align}
 \sigma=\frac{s^3}{160\pi F^4},
\label{s3}
\end{align}
presented by a black line in Fig.~\ref{fig:xsec_rs}. The contribution
without the four-point amplitude is also shown as a reference, where one 
can see the effect of the huge cancellation. It should be noted here
that the low-energy limit, which is always assumed in the previous
studies~\cite{Nachtmann:1984xu,Brignole:1997sk,Checchia:1999dr,Gopalakrishna:2001iv}, 
may not be a good approximation for future colliders since the selectron
masses should be less or of the order of the SUSY breaking scale and
might be within the reach of the CM energies. Therefore, one should
consider the full expression of the cross
section. Figure~\ref{fig:xsec_rs} indeed shows that, as $\sqrt{s}$ is  
increasing, the effect of the selectron mass becomes significant. When
the CM energy is bigger than the selectron mass, $\sqrt{s}>m_{\se}$, the
contribution from $\mathcal{M}^c$ becomes more important than that from
$\mathcal{M}^{t,u}$. We note that the current gravitino mass bound by
the $\gld\gld(+\gamma)$ production could weaken if the selectrons are
light enough.   

Finally, we briefly discuss the unitarity bound. The projected partial
wave amplitude is given by 
\begin{align}
  {\cal J}^J_{\lam_1,\lam_3}
 =\frac{1}{32\pi}\int_{-1}^1d \cos{\theta}\,
  d^J_{\lam_1\lam_3}(\theta)\,\mathcal{M}^{}_{\lam_1,\lam_3}
\end{align}
with the Wigner $d$-function. Unitarity requires the lowest
non-vanishing partial wave to be 
$|{\cal J}^{J=1}_{\lam_1,\lam_3}|<1/2$, leading to the upper bound of
the cross section, which is shown by a gray line in
Fig.~\ref{fig:xsec_rs}. One can see that the lighter selectrons remedy
the bad unitarity behavior.  
It should also be noted that, since we consider the effective model
which is valid up to $m_{\rm SUSY}/F$, a higher energy requires a 
higher SUSY breaking scale (i.e. a heavier gravitino) or lighter SUSY
particles for reliable predictions.

\subsection{Neutralino-gravitino associated production}

Gravitino production in association with a neutralino and the subsequent
neutralino decay, 
\begin{align}
 e^+e^-\to\nored\gld\to\gamma\gld\gld,
\end{align}
leads to the $\gamma+\Emiss$ signal already at the leading
order~\cite{Fayet:1986zc,Dicus:1990vm,Lopez:1996gd,Lopez:1996ey,Baek:2002np,Mawatari:2011cu}.%
\footnote{The monophoton signal of $\nored\gld$ production via the Higgs
decay at the LHC was studied in~\cite{Petersson:2012dp}.}  
We refer to the recent study~\cite{Mawatari:2011cu} for a detailed
discussion. 

\begin{figure*}
\center
 \includegraphics[width=0.88\columnwidth]{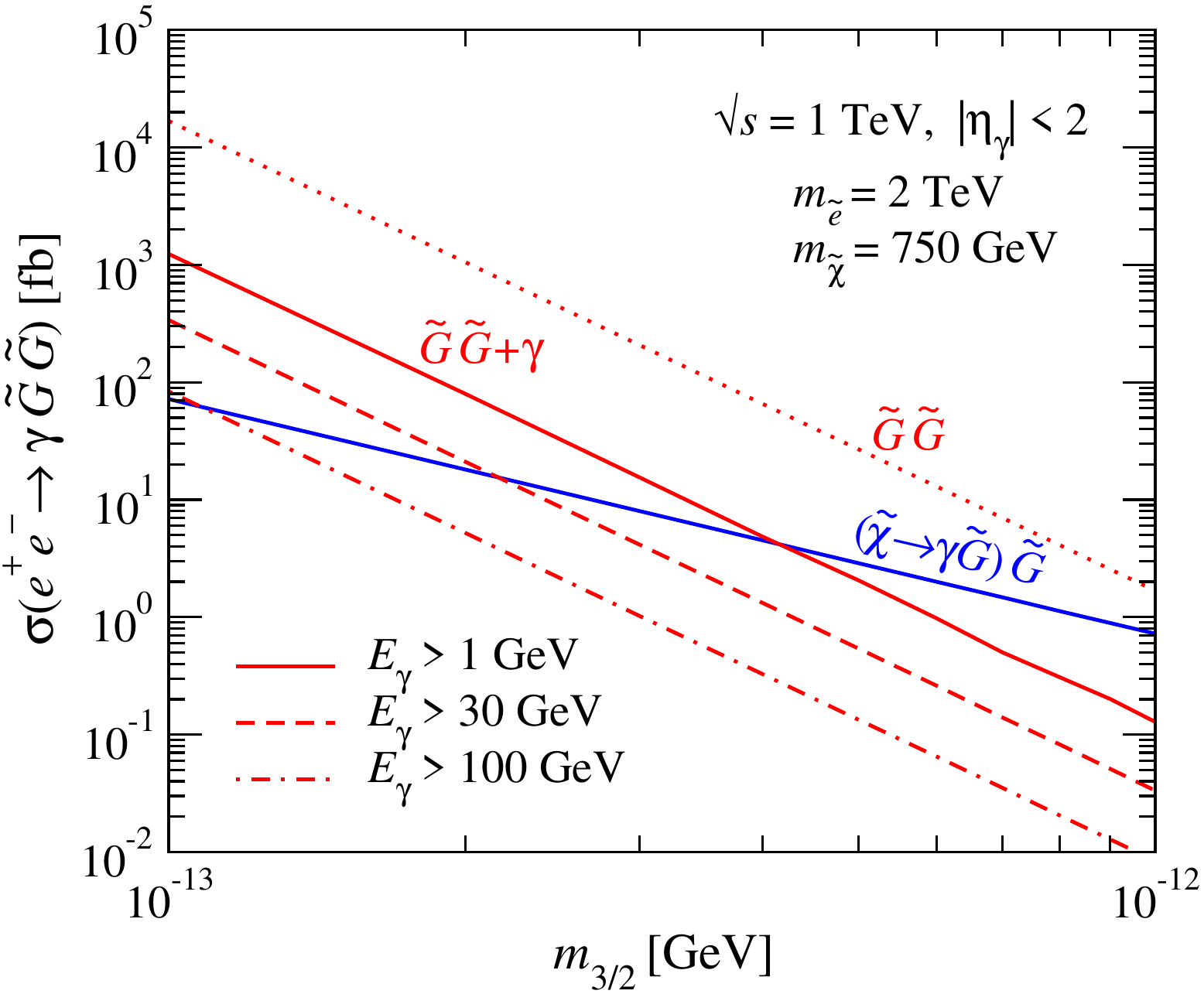}\quad
 \includegraphics[width=0.88\columnwidth]{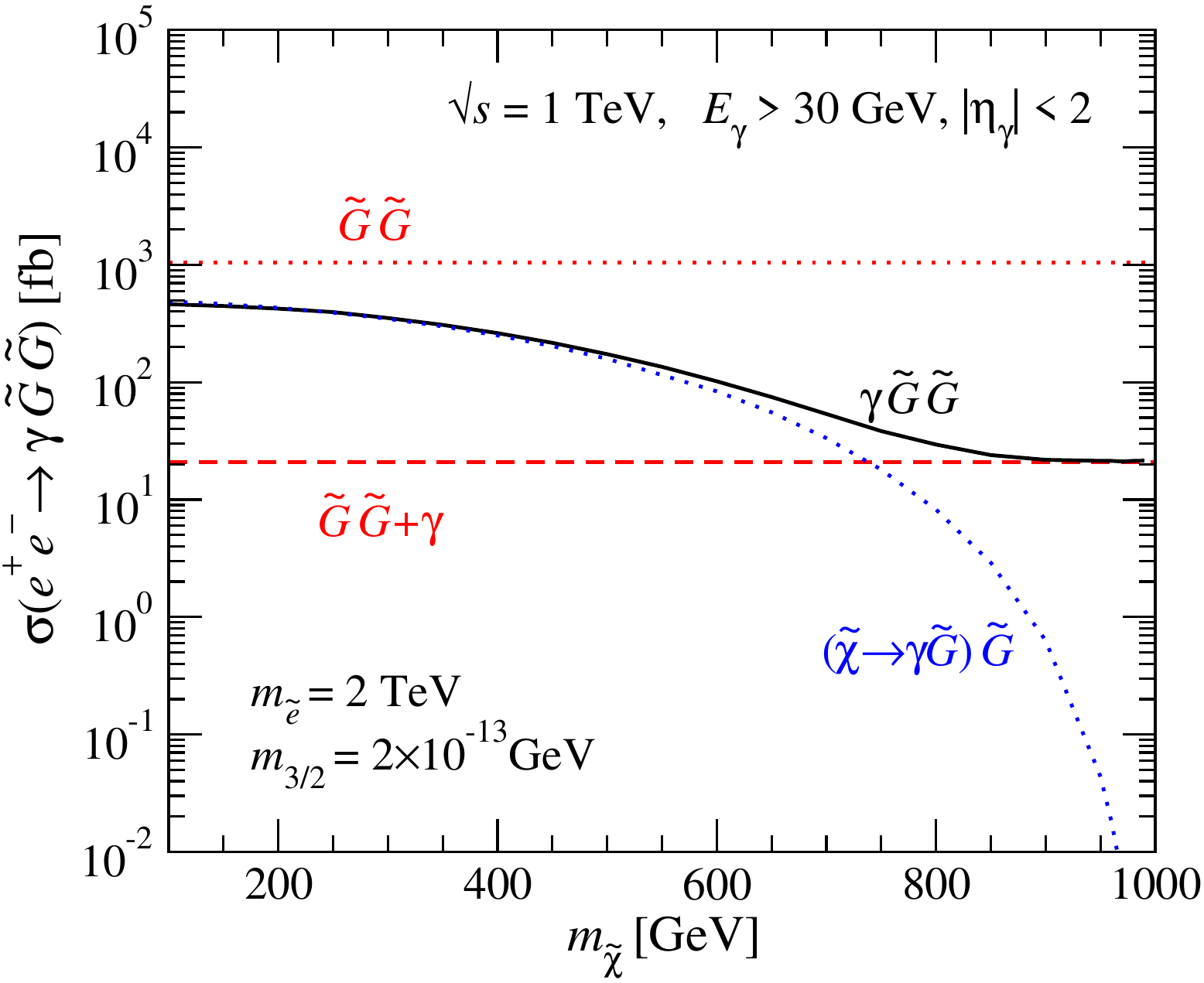}
 \caption{Total cross sections of $e^+e^-\to\gamma\gld\gld$ as a
 function of the gravitino mass (left) and the neutralino mass (right)
 for $m_{\se_{\pm}}=2$~TeV at $\sqrt{s}=1$~TeV. The contributions of the
 gravitino pair production and the neutralino-gravitino associated
 production are separately shown by red and blue lines,
 respectively. The cross section of $e^+e^-\to\gld\gld$ is also shown by
 a red dotted line as a reference. On the left plot the contributions of
 the $\gld\gld$ production are shown with different photon energy cuts
 $E_{\gamma}>1$, 30 and 100~GeV, while the $E_{\gamma}$ cut is fixed at
 30~GeV on the right.} 
\label{fig:mgld_dep} 
\end{figure*}

Here, we briefly point out two important features of this process.
First, unlike the gravitino pair production~\eqref{sig_gldgld}, 
the total cross section is inversely proportional to the square of the
gravitino mass
\begin{align}
 \sigma(\nored\gld)\propto 1/m_{3/2}^2,
\label{sig_n1gld}
\end{align}
as seen in the left plot in Fig.~\ref{fig:mgld_dep}, and hence the
sensitivity to the gravitino mass is weaker than in the $\gld\gld$
production. The cross section depends also on the $t,u$-channel exchange
selectron masses, and increases for the heavier selectrons as in the
$\gld\gld$ production. 

\begin{figure*}
\center
 \includegraphics[width=0.32\columnwidth]{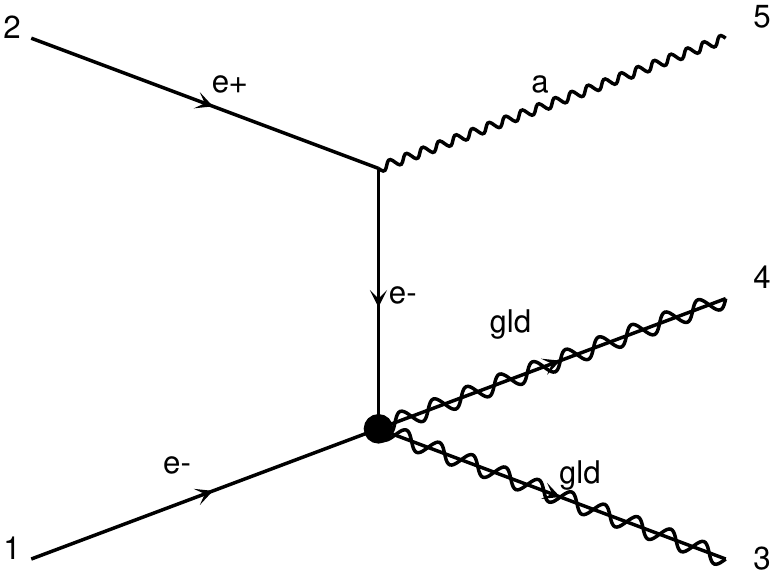}
 \includegraphics[width=0.32\columnwidth]{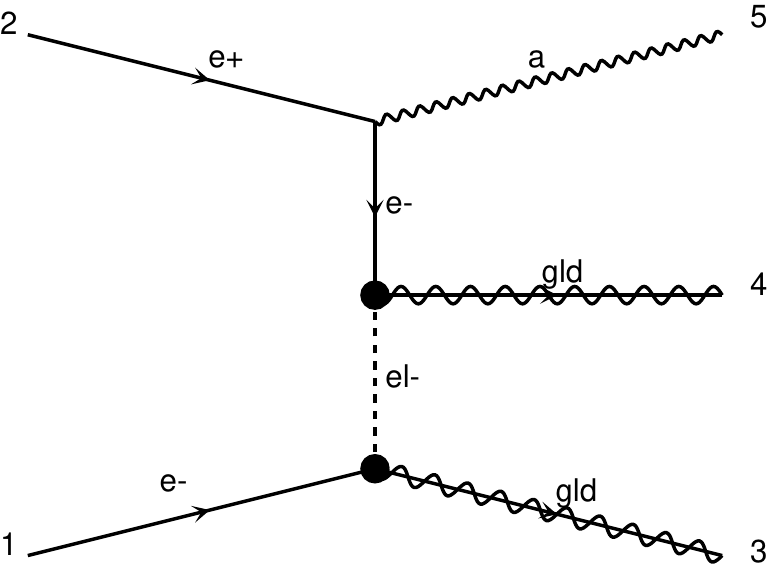}
 \includegraphics[width=0.32\columnwidth]{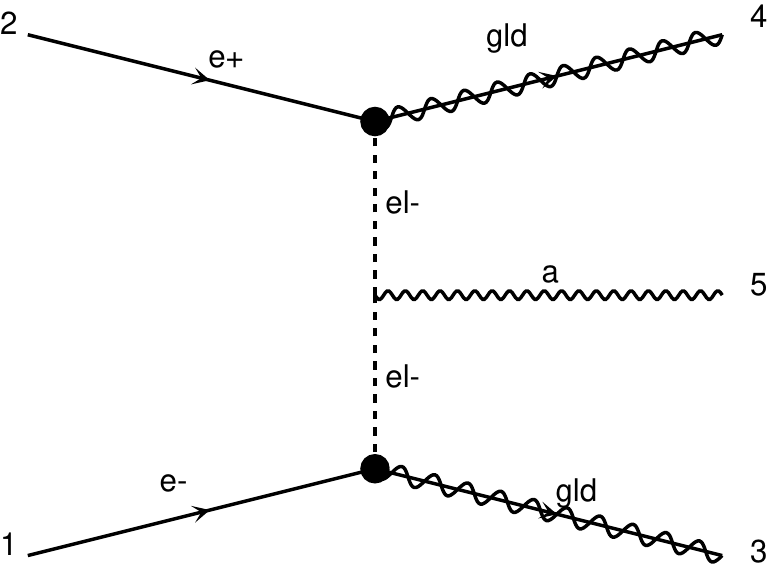}
 \includegraphics[width=0.32\columnwidth]{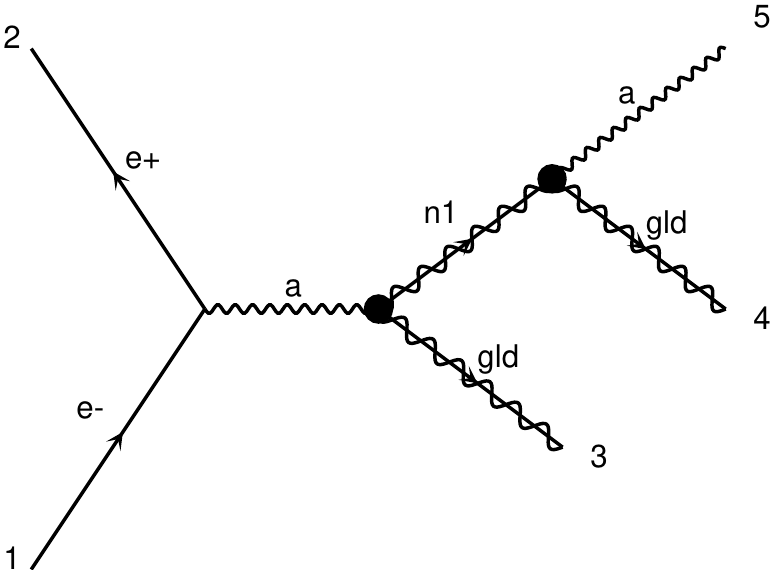}
 \includegraphics[width=0.32\columnwidth]{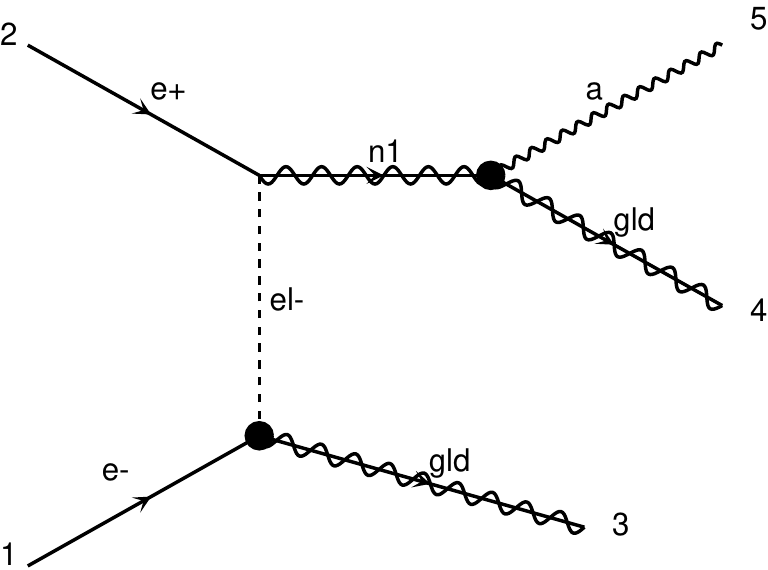}
 \includegraphics[width=0.32\columnwidth]{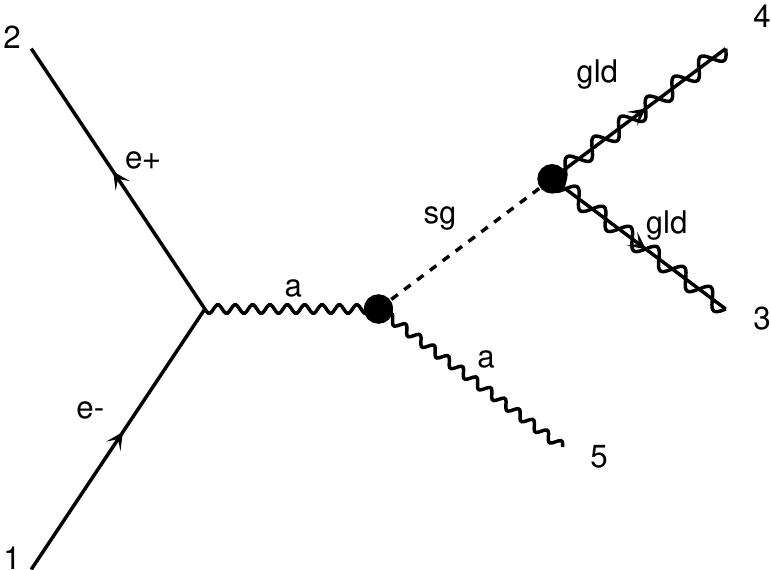}
 \caption{Representative Feynman diagrams for 
 $e^+e^-\to\gld\gld\gamma$, generated by (modified) 
 {\sc MadGraph 5}~\cite{Alwall:2011uj}. {\tt n1} and
 {\tt sg} denote a neutralino and a sgoldstino, respectively.}
\label{fig:diag}
\end{figure*} 

Second, since the $\nored\to\gamma\gld$ decay is isotropic, the photon
distribution is given by purely kinematical effects of the decaying
neutralino. The partial decay width for a photino-like neutralino is
given by 
\begin{align}
 \Gamma(\nored\to\gamma\gld)=\frac{m^5_{\nored}}{16\pi F^2}.
\end{align}
For instance, for $m_{\nored}=750$~GeV and 
$m_{3/2}=2\times10^{-13}$~GeV (i.e. $\sqrt{F}\approx 918$~GeV), the
width is 6.6~GeV. With the neutralino being the NLSP, the branching
ratio is unity, $B(\nored\to\gamma\gld)=1$.

\subsection{Physics parameters}

To examine a viable SUSY parameter space for the $\gamma+\Emiss$ signal
at future $e^+e^-$ colliders, we present in Fig.~\ref{fig:mgld_dep}
the total cross sections of $e^+e^-\to\gamma\gld\gld$ at
$\sqrt{s}=1$~TeV as a function of the gravitino mass (left) and the
neutralino mass (right), where we fix the left- and right-handed
selectron masses at 2~TeV. 
The representative Feynman diagrams for the process are depicted in
Fig.~\ref{fig:diag}.
The contributions of the $\gld\gld$ and
$\nored\gld$ productions are separately shown by red and blue lines,
respectively. 

As discussed in~\eqref{sig_gldgld} and \eqref{sig_n1gld} and shown in
the left plot in Fig.~\ref{fig:mgld_dep}, the cross sections of the both
subprocesses strongly depend on the gravitino mass. 

The monophoton signal from the gravitino pair ($\gld\gld+\gamma$) is
suppressed by the QED coupling $\alpha$ with respect to the two-to-two
process and strongly depends on the kinematical cuts due to the soft and
collinear singularity of the initial state radiation. The cut dependence
on the photon energy is presented in the left plot in
Fig.~\ref{fig:mgld_dep}. On the other hand, since the energy of the 
photons coming from the neutralino decay is restricted as  
\begin{align}
 \frac{m^2_{\nored}}{2\sqrt{s}}<E_{\gamma}<\frac{\sqrt{s}}{2},
\label{photon_energy}
\end{align}
the signal of $\nored\gld$ is not affected by the lower cuts on the
photon energy unless the neutralino is light. 

In the following, we impose the minimal cuts for the detection of
photons as
\begin{align}
 E_{\gamma}>0.03\,\sqrt{s},\quad |\eta_{\gamma}|<2,
\label{minimal_cuts}
\end{align}
and fix the gravitino mass at $2\times 10^{-13}$~GeV, which lies above
the current exclusion limit by the jet$+\Emiss$ search at the LHC for
the gravitino production in association with a gluino or a squark with
masses around 500~GeV~\cite{ATLAS:2012zim}.%
\footnote{Astrophysics observables, e.g. energy losses of red giant
stars~\cite{Fukugita:1982eq} and supernova~\cite{Dicus:1997sw} can also
provide the lower limit on the gravitino mass. But their limits are less 
stringent.}
\footnote{As discussed in Sect.~\ref{sec:grav_pair_prod}, reliability of 
the effective theory calculation can also constrain the model parameter
space.} 

The right plot of Fig.~\ref{fig:mgld_dep} shows the neutralino mass
dependence of the full signal cross section with the minimal
cuts~\eqref{minimal_cuts}. While the $\gld\gld$ contribution is
independent of the neutralino mass, the contribution from the
$\nored\gld$ production is strongly suppressed when the neutralino mass
approaches the CM energy due to the phase space closure. Therefore, the
dominant subprocess can be different for different neutralino
masses, giving rise to distinctive photon spectra. It should be noted
that the interference between the two subprocesses is very small unless
the neutralino width is too large. We verified this numerically by
computing the two subprocess separately and checking that the sum of
those reproduces the full $e^+e^-\to\gamma\gld\gld$ cross section, as in
the figure. 
We suppress a possible contribution from the sgoldstinos
by taking their masses to be
too heavy to be produced on-shell.%  
\footnote{We note that sgoldstinos with masses much smaller than the
selectron mass do not obey a naturalness
criterion~\cite{Brignole:1998uu}.} 
We note that, if those are lighter than the $e^+e^-$ collision energy, 
the sgoldstino production in association with a photon and the
subsequent decay contributes to the $\gamma\gld\gld$ final state.
In Appendix~\ref{sec:aa} we briefly discuss the effect of sgoldstinos
in the $\gamma\gamma\to\gld\gld$ process.

In the following, we focus on three different neutralino masses which
exemplify different distributions. First, we fix the neutralino mass
at 750 GeV so that $\sigma(\nored\gld)\sim\sigma(\gld\gld+\gamma)$. We
subsequently take a lighter (heavier) neutralino at 650 (850)~GeV so
that the $\nored\gld$ ($\gld\gld$) production is dominant.

\subsection{Technical setup and validation}

Before moving to the simulation, let us comment on our model
implementation and the validation. As mentioned in
Sect.~\ref{sec:intro}, the current {\sc MadGraph~5}
(v2.0.2)~\cite{Alwall:2011uj} does not support four-fermion vertices
involving more than one Majorana particle, and hence does not accept our
{\sc UFO} model file~\cite{Degrande:2011ua,deAquino:2011ub} generated
with {\sc FeynRules}~\cite{Alloul:2013bka}. Therefore, first, we
modified {\sc MadGraph~5} to allow us to import the model. Second, after  
generating the process, the corresponding four-point contact amplitudes
should be modified by hand to have correct fermion flows. We have
explicitly checked our numerical results of the total and differential
cross sections by comparing with the analytic results for the two-to-two
process in Sect.~\ref{sec:grav_pair_prod} as well as for the
two-to-three process in the low-energy limit, 
$\sqrt{s}\ll m_{\se,\nored,S,P}$, given in~\cite{Brignole:1997sk}. We
have also checked precise agreements for the $\nored\gld$ process with
the previous model
implementations~\cite{Mawatari:2011jy,Argurio:2011gu,Mawatari:2012ui},
which are constructed based on the effective gravitino Lagrangian in
terms of the component fields, i.e. not by using the superspace
module. We note that our model implementation allows us to generate
different contributing processes, i.e. $\gld\gld$ and $\nored\gld$,
within one event simulation.

%%%%%%%%%%%%%% Begin Section 4 %%%%%%%%%%%%%%%%%%%%%%%%%%%%%%%%%%%%%%%%% 
\section{Monophoton plus missing energy}\label{sec:signal}

We now perform the simulation of monophoton events with missing energy
for a future $e^+e^-$ collider. An irreducible SM background comes from 
$e^+e^-\to\gamma\nu\bar{\nu}$. To remove contributions from
$e^+e^-\to\gamma Z\to\gamma\nu\bar{\nu}$, we impose the $Z$-peak cut 
\begin{align}
 E_{\gamma}<\frac{s-m_Z^2}{2\sqrt{s}}-5\Gamma_Z,
\label{Zpeak_cut}
\end{align}
in addition to the minimal cuts~\eqref{minimal_cuts}. The background
from the $t$-channel $W$-exchange process, which is the most significant
one, can be efficiently reduced by using a positively polarized $e^-$
beam and a negatively polarized $e^+$ beam. 

In Table~\ref{tb:cross_section}, the signal cross sections of each
subprocess, $\nored\gld$ and $\gld\gld$, as well as the SM background
at $\sqrt{s}=1$~TeV are presented without and with polarized $e^{\pm}$
beams, where we take the beam polarization
$P_{e^{\pm}}\, (|P_{e^{\pm}}|\leq1)$ as%
\footnote{$|P_{e^-}|>0.8$ and $|P_{e^+}|>0.5$ are designed at the
International Linear Collider (ILC)~\cite{BrauJames:2007aa}.}
\begin{align} 
 (P_{e^-},P_{e^+})=(0.9,-0.6),
\label{polarization}
\end{align}
and apply the kinematical cuts of~\eqref{minimal_cuts} and
\eqref{Zpeak_cut}. For the SUSY signal, we take the three benchmark
neutralino masses with the gravitino mass fixed at
$2\times10^{-13}$~GeV for $m_{\se}=1$ and 2~TeV. 
As discussed in the previous section, heavier selectrons give the higher 
cross sections of the both subprocesses.   
Since the signal cross section with $e^{\pm}$ beam polarizations is
given by 
\begin{align}
 \sigma(P_{e^-},P_{e^+}) &= 2\sum_{\lam_1}
 \Big(\frac{1+P_{e^-}\lam_1}{2}\Big)\Big(\frac{1-P_{e^+}\lam_1}{2}\Big)\,
 \sig_{\lam_1},
\end{align}
the signal cross sections are enhanced by a factor of 1.54 with the
above polarizations. 
On the other hand, the SM background is significantly reduced. 

\begin{figure}
\center
 \includegraphics[width=0.492\columnwidth]{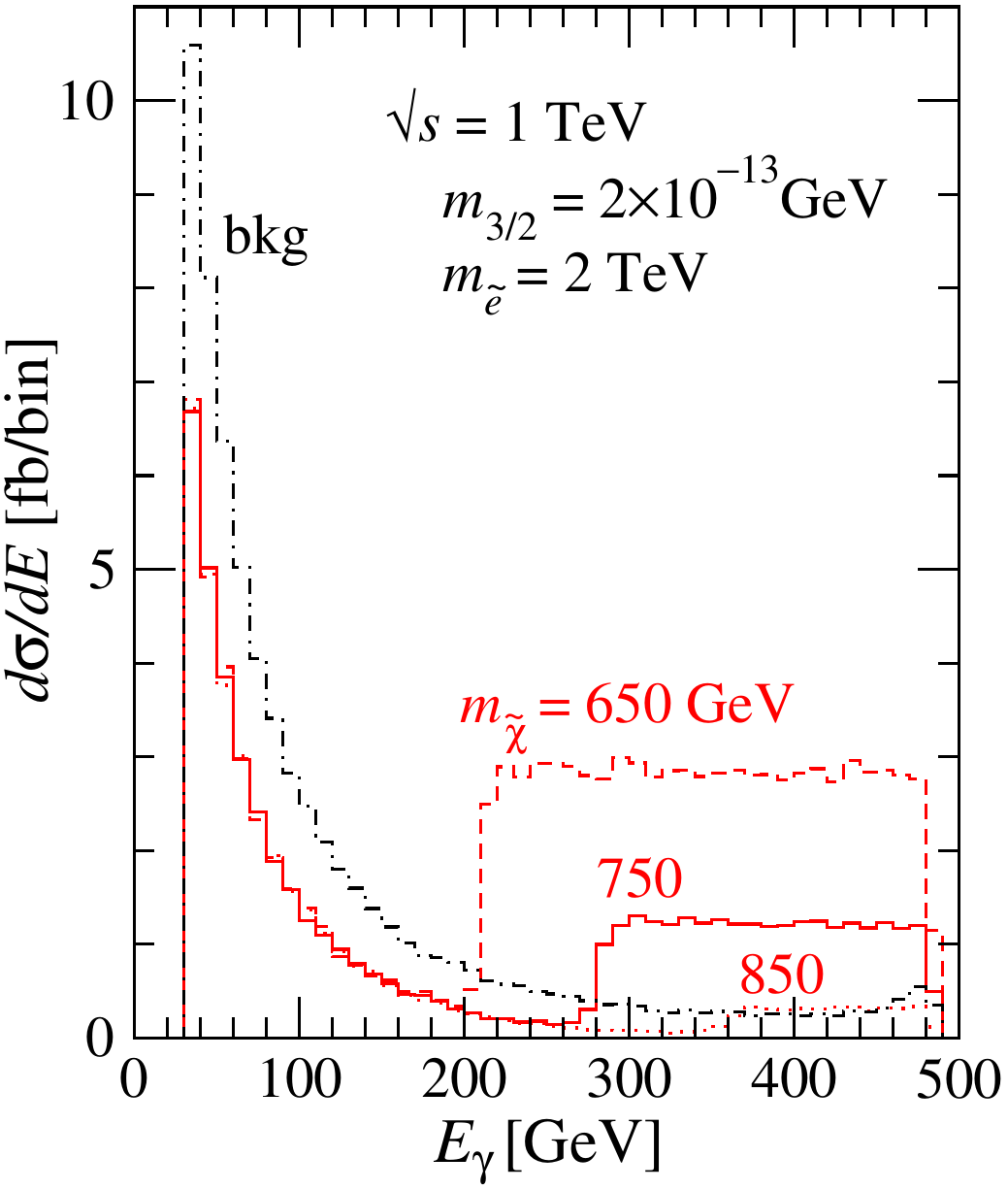}
 \includegraphics[width=0.492\columnwidth]{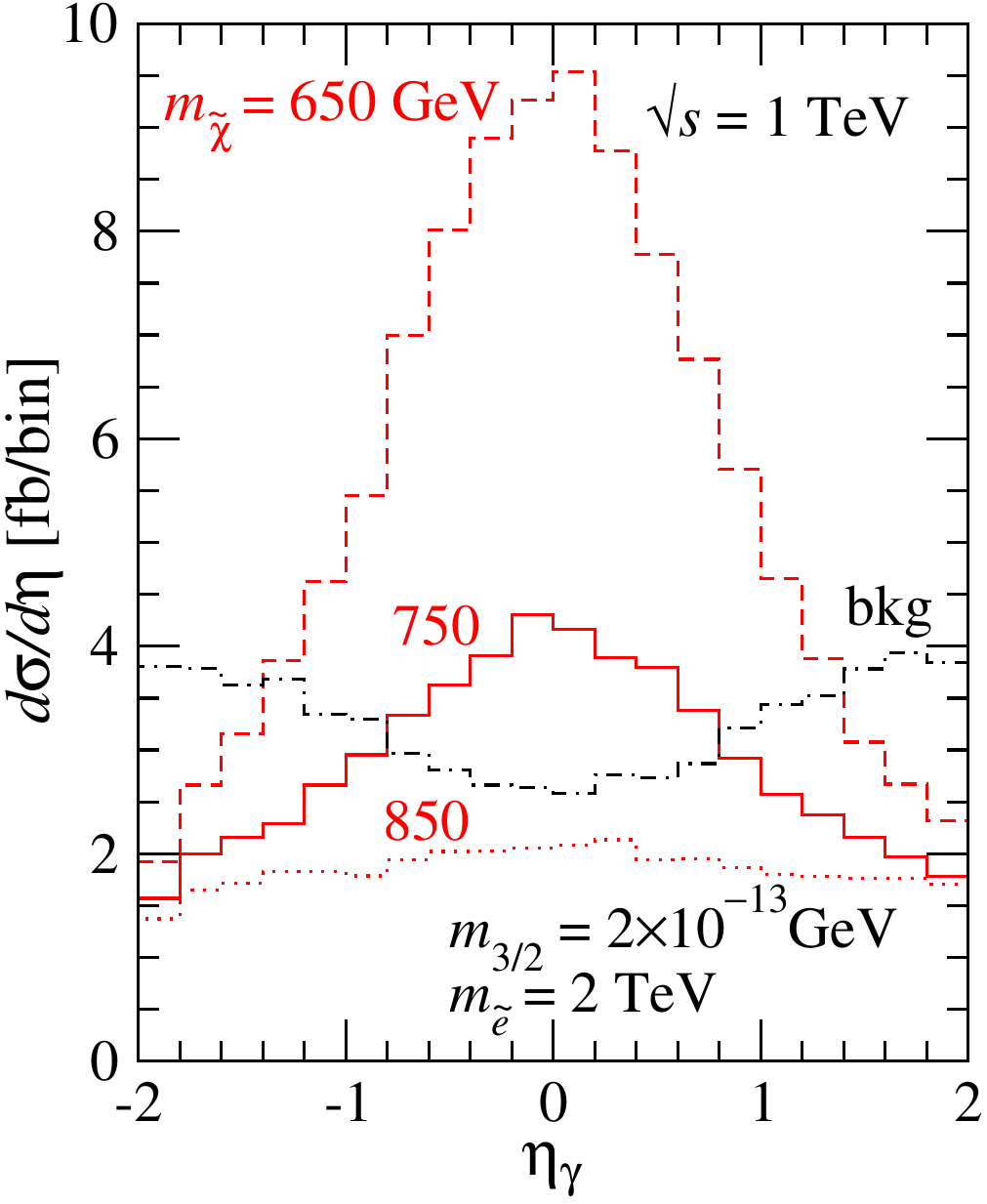}
\caption{Photon energy (left) and rapidity (right) distributions for
 $e^+e^-\to\gamma\gld\gld$ at $\sqrt{s}=1$~TeV for different neutralino
 masses with $m_{3/2}=2\times10^{-13}$~GeV and
 $m_{\se_{\pm}}=2$~TeV. The kinematical cuts in~\eqref{minimal_cuts} and
 \eqref{Zpeak_cut} as well as the beam polarizations 
 in~\eqref{polarization} are applied. The SM background is also shown.} 
\label{fig:photon_energy} 
\end{figure}

Figure~\ref{fig:photon_energy} presents the photon energy $E_{\gamma}$
(left) and rapidity $\eta_{\gamma}$ (right) distributions for the three
signal benchmarks and for the SM background. The signal energy spectra
show two distinct features. First, there is a peak in the low-energy
region which arises from the $\gld\gld$ production process since the
initial state radiation is dominant as in the SM background. We also
note that the low-energy spectra are independent of the neutralino
mass. Second, there is a flat contribution in the high-energy region
coming from $\nored\gld$ production, reflecting the isotropic neutralino
decay. The contribution becomes smaller for the heavier neutralino (see
also Table~\ref{tb:cross_section}), and the lower edge allows us to
extract the neutralino mass from~\eqref{photon_energy}. 

The rapidity distributions are distinctive between the signal and the SM
background. The photon coming from $\gld\gld$ production gives a flat
$\eta_{\gamma}$ distribution while the photon coming from the neutralino
decay results in the central region (see~\cite{Mawatari:2011cu} for a
detailed discussion of the selectron mass dependence). In contrast, the
photons of the SM background are emitted in the forward region. 

\begin{table*}
\center
 \caption{Cross sections in fb unit of each subprocess for the signal
 $e^+e^-\to\gamma\gld\gld$ and of the SM background
 $e^+e^-\to\gamma\nu\bar{\nu}$ at $\sqrt{s}=1$~TeV, without and with
 beam polarizations. The kinematical cuts of~\eqref{minimal_cuts} and
 \eqref{Zpeak_cut} are applied. For the signal three (two) different
 neutralino (selectron) masses are taken with the gravitino mass fixed
 at $2\times10^{-13}$~GeV.}  
\begin{tabular}{rr||rr|rr|rr}
\hline
 && \multicolumn{2}{|c|}{($m_{\se}=1$~TeV)}
  & \multicolumn{2}{|c|}{($m_{\se}=2$~TeV)} && [fb]\\
 $(P_{e^-},P_{e^+})$&$m_{\nored}$ [GeV]
 &$\nored\gld$& $\gld\gld$&$\nored\gld$& $\gld\gld$
 &SM bkg & \\\hline\\[-3.5mm]
 &650  &19.7 &&49.2 &&\\
 $(0,0)$ &750 &6.0 &10.4 &15.8 &21.1&1452\\
 &850 &1.0 & &2.5 &&\\\hline\\[-3.5mm]
 &650  &30.4 & &75.8  &&\\
 $(0.9,-0.6)$ &750 &9.2 &16.1 &24.3 &32.7&64.9\\
 &850  &1.5 & &3.4&& \\
\hline
\end{tabular}
\label{tb:cross_section}
\end{table*}

\begin{figure}
\center
 \includegraphics[width=0.7\columnwidth]{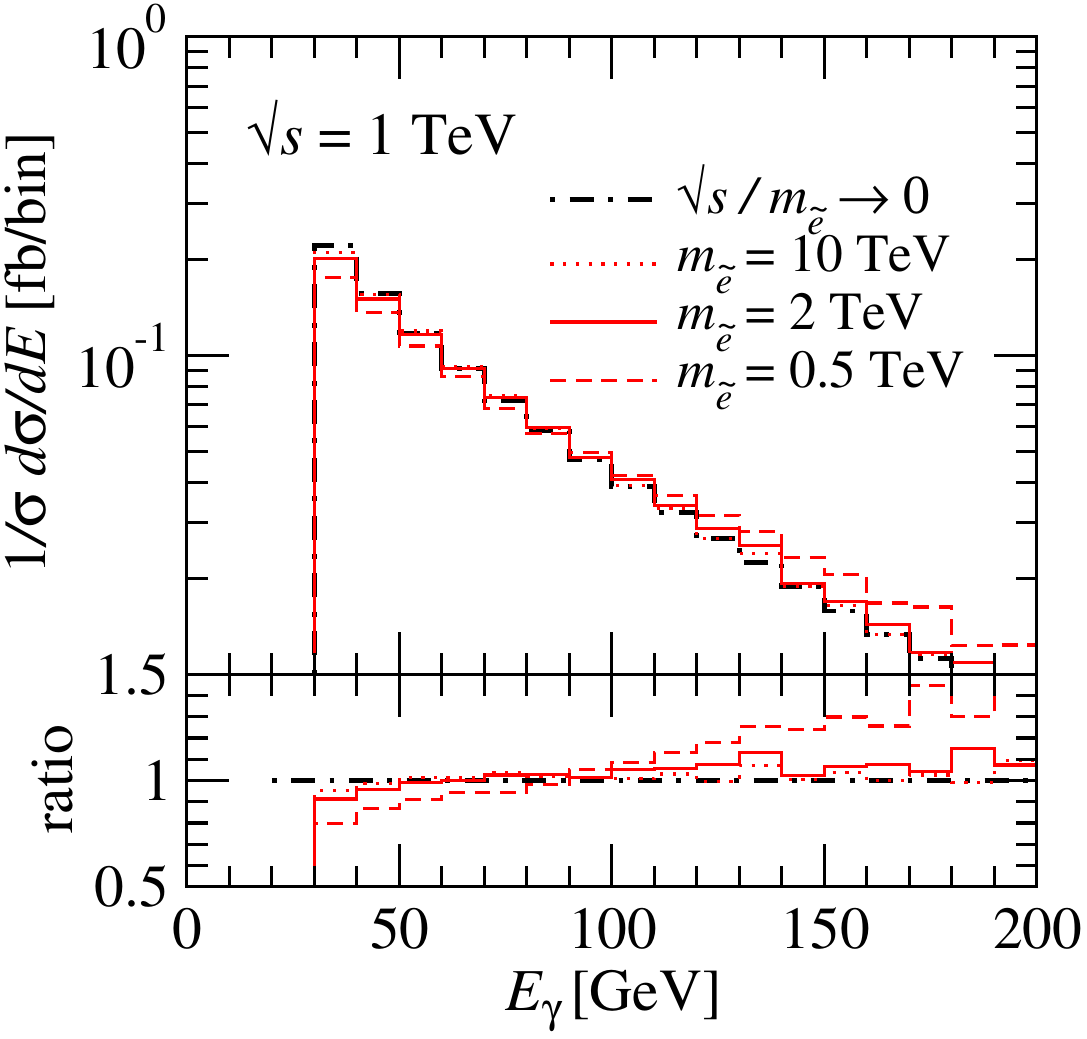}
 \caption{Normalized photon energy distributions for 
 $e^+e^-\to\gamma\gld\gld$ at $\sqrt{s}=1$~TeV for $m_{\se_{\pm}}=0.5$,
 2, 10~TeV and for the high-mass limit, where the kinematical  
 cuts~\eqref{minimal_cuts} are applied. The ratios to the case in the
 high-mass limit are also shown.}   
\label{fig:ratio_norm_xs} 
\end{figure}

Finally, we discuss the selectron mass dependence of the low-energy
peak, which arises purely from $\gld\gld$ production. As discussed in
Sect.~\ref{sec:grav_pair_prod}, the total rate of the
$e^+e^-\to\gld\gld$ process depends on the selectron masses. In
addition, the photon spectrum becomes harder for lighter selectrons;
see Fig.~\ref{fig:ratio_norm_xs}, where we show the normalized photon
energy distributions for $m_{\se_{\pm}}=0.5$, 2, 10~TeV and for the 
$\sqrt{s}/{m_{\se}}=0$ limit~\cite{Brignole:1997sk}. The  distribution
for $m_{\se_{\pm}}=10$~TeV is in good agreement with the one in the
high-mass limit. We note that in this limit the
$e^+e^-\to\gamma\gld\gld$ differential cross section can be described by
the $e^+e^-\to\gld\gld$ cross section times the standard photon
splitting function in a good approximation~\cite{Brignole:1997sk}.

%%%%%%%%%%%%%% Begin Section 5 %%%%%%%%%%%%%%%%%%%%%%%%%%%%%%%%%%%%%%%%% 
\section{Summary}\label{sec:summary}

Direct gravitino productions can be observed in current and future
collider experiments if the gravitino is very light. In this article, we  
revisited gravitino pair production and neutralino-gravitino associated
production, and studied the $\gamma+\Emiss$ signal for future $e^+e^-$
colliders. 

By using the superspace formalism, we constructed a simple SUSY QED
model that allows us to study the parameter space for the both
processes, and implemented the model in the {\sc FeynRules} and
{\sc MadGraph~5} frameworks. We note that special implementations are 
needed to treat the Majorana four-fermion interaction in
{\sc MadGraph~5}.  

We discussed the parameter dependence of the signal cross sections in
detail, and showed that the relative importance between the two signal
processes varies with the gravitino and neutralino masses as well as
with kinematical cuts. 

We performed the event simulation for the SUSY signal as well as the SM
background, taking into account the signal selection cut and the beam
polarizations, and showed that the photon spectra from the two
subprocesses are very distinctive. This is because the photon coming
from the $\gld\gld$ production is mostly initial state radiation, while
the $\nored\gld$ associated production process leads to an energetic 
photon from the neutralino decay. We expect that future $e^+e^-$
colliders could explore the parameter space around our benchmark points
and hence provide information on the masses of the relevant SUSY
particles as well as the SUSY breaking scale. 

Before closing, we note that the extension of our simple SUSY QED model
to the general MSSM is straightforward, which is applicable for hadron
colliders.

%AAAAAAAAAAAAAAAAAAAAAAAAA
\begin{small}
\section*{Acknowledgments}

We wish to thank B.~Fuks and O.~Mattelaer for their help with
{\sc FeynRules} and {\sc MadGraph5} and F.~Maltoni and 
C.~Petersson for useful discussions. 
We also thank Y.~Takaesu and P.~Tziveloglou for helpful discussions and
comments on the draft.  
This work has been supported in part 
by the Belgian Federal Science Policy Office through the Interuniversity
Attraction Pole P7/37,
and
by the Strategic Research Program ``High Energy Physics'' and the
Research Council of the Vrije Universiteit Brussel.   
\end{small}

\appendix
%%%%%%%%%%%%%% Begin App. 1 %%%%%%%%%%%%%%%%%%%%%%%%%%%%%%%%%%%%%%%%
\section{Lagrangian in terms of the component fields}\label{sec:lag}

In Sect.~\ref{sec:model} we gave the Lagrangian of our model in terms of 
the superfields. In this appendix, for completeness, we present the
corresponding interaction Lagrangian in terms of the component
fields. The relevant terms of the effective interaction Lagrangian among
gravitinos (i.e. goldstinos) $\psi_{\gld}$ and fields in the visible
sector, that is, right- and left-handed selectron
$\phi_{\se_{\pm}}$, electron $\psi_e$, photino-like neutralino
$\psi_{\nored}$,%
\footnote{See e.g. Appendix~A in~\cite{Mawatari:2012ui} for the general
case of the neutralino mixing.}
and photon $A^{\mu}$ are given in the four-component notation by  
\begin{align}
\mathcal{L}_{\gld}&=\mp\frac{im_{\se_{\pm}}^2}{F}(\psibar_{\gld}P_{\pm}\psi_e\phi^{*}_{\se_{\pm}}-\psibar_eP_{\mp}\psi_{\gld}\phi_{\se_{\pm}})\nn \\
&\quad-\frac{m_{\nored}}{4\sqrt{2}F}\,\psibar_{\gld}[\gamma^{\mu},\gamma^{\nu}]\psi_{\nored}F_{\mu\nu}\nn \\
&\quad-\frac{m^2_{\se_{\pm}}}{F^2}\,\psibar_eP_{\mp}\psi_{\gld}\,\psibar_{\gld}P_{\pm}\psi_e, 
\end{align}
where $P_{\pm}=\frac{1}{2}(1\pm\gamma^5)$ is the chiral projection
operator and $F_{\mu\nu}=\partial_{\mu}A_{\nu}-\partial_{\nu}A_{\mu}$
the photon field strength tensor. The interactions among sgoldstino
$\phi=\frac{1}{\sqrt{2}}(\phi_S+i\phi_P)$ and gravitino or photon
are given by
\begin{align}
\mathcal{L}_{S,P}&=
-\frac{m^2_{\phi}}{2\sqrt{2}F}\,\psibar_{\gld}(\phi_S+i\gamma^5\phi_P)\psi_{\gld}\nn\\
&\quad+\frac{m_{\nored}}{2\sqrt{2}F}(\phi_SF^{\mu\nu}F_{\mu\nu}-\phi_PF^{\mu\nu}\tilde{F}_{\mu\nu}),
\end{align}
where 
$\tilde{F}_{\mu\nu}=\frac{1}{2}\epsilon_{\mu\nu\alpha\beta}F^{\alpha\beta}$
is the dual tensor with  $\epsilon_{0123}=+1$. All other relevant terms
in the visible sector are
\begin{align}
\mathcal{L}_{\rm vis}&=g_e\psibar_e\gamma_{\mu}\psi_eA^{\mu} 
 +ig_e(\phi^{*}_{\se_{\pm}}\overleftrightarrow{\partial_{\mu}}\phi_{\se_{\pm}})A^{\mu}
 \nn \\
 &\quad\mp\sqrt{2}g_e(\psibar_{\nored}P_{\pm}\psi_e\phi^*_{\se_{\pm}} +
 \psibar_eP_{\mp}\psi_{\nored}\phi_{\se_{\pm}}),
\end{align}
where $g_e=\sqrt{4\pi\alpha}$ is the QED coupling constant.

We note that we follow the convention of the SUSY Les Houches
accord~\cite{Skands:2003cj} for the covariant derivative and the
gaugino and gravitino field definitions. To translate our Lagrangian
into the {\sc FeynRules} convention, one has to change the coupling as
$g_e\to-g_e$, and redefine the fields as 
$\psi_{\nored}\to -\psi_{\nored}$ and $\psi_{\gld}\to -\psi_{\gld}$.

\section{Gravitino pair production in $\gamma \gamma$ collisions}
\label{sec:aa}

In this article we assumed that the sgoldstinos are too heavy to be
produced on-shell, and hence those are irrelevant to the 
$e^+e^-\to\gamma\gld\gld$ process. However, our model has no limitation 
to study processes involving sgoldstinos by construction in the
superspace formalism. In this appendix, to validate our model
implementation of sgoldstinos, we discuss gravitino pair production in
$\gamma\gamma$ collisions, where the sgoldstinos play an important role
for the unitarity~\cite{Bhattacharya:1988ey,Bhattacharya:1988zp}.%
\footnote{The amplitudes were calculated by using the explicit 
 spin-3/2 gravitino wavefunction analytically in the high-energy limit 
 in~\cite{Bhattacharya:1988ey,Bhattacharya:1988zp} and numerically 
 in~\cite{Christensen:2013aua}, including the spin-2 graviton exchange
 diagram.} 

\begin{figure}
\center
 \includegraphics[width=1\columnwidth]{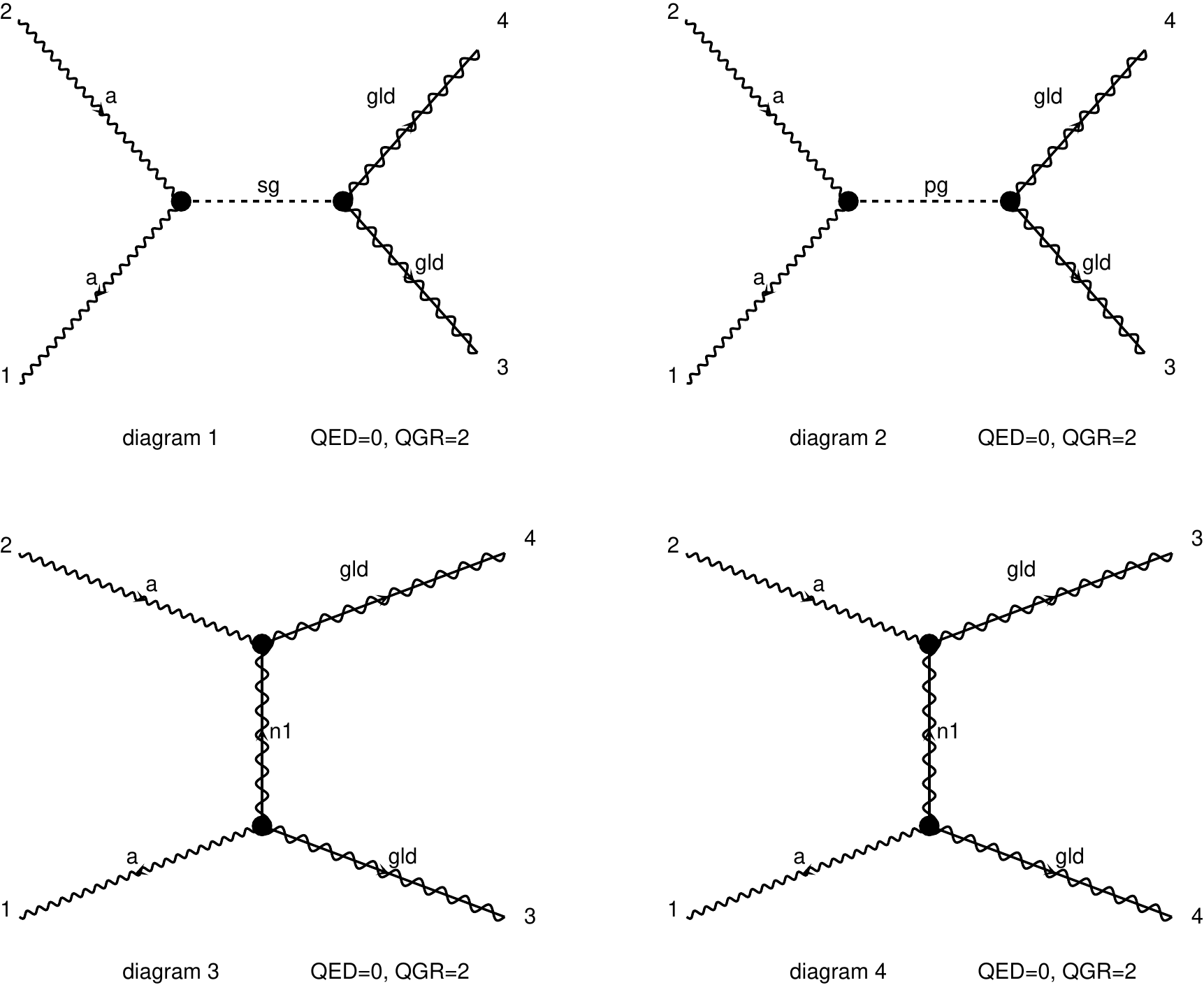}
 \caption{Feynman diagrams for gravitino pair production in
 $\gamma\gamma$ collisions, generated by 
 {\sc MadGraph 5}~\cite{Alwall:2011uj}. {\tt gld}, {\tt sg}, {\tt pg}, 
 and {\tt n1} denote a gravitino, a scalar sgoldstino, a pseudoscalar
 sgoldstino, and a neutralino, respectively.}
\label{fig:diagramaa}
\end{figure}

Similar to Sect.~\ref{sec:grav_pair_prod}, we present the helicity
amplitude explicitly for the process
\begin{align}
 \gamma\left(p_1,\lambda_1\right) + \gamma\left(p_2,\lambda_2\right)
 \to 
  \gld\Big(p_3,\frac{\lambda_3}{2}\Big) 
 +\gld\Big(p_4,\frac{\lambda_4}{2}\Big),
\label{xs_aa_gldgld} 
\end{align}
where the four momenta ($p_i$) and helicities ($\lambda_i=\pm1$) are
defined in the center-of-mass (CM) frame of the $\gamma\gamma$
collision. As seen in Fig.~\ref{fig:diagramaa}, in our SUSY QED model, 
the helicity amplitudes are given by the sum of the $s$-channel scalar
($S$) and pseudoscalar ($P$) sgoldstino amplitudes and the $t,u$-channel  
photino-like neutralino exchange amplitudes:
\begin{align}
 &\mathcal{M}_{\lambda_1\lambda_2,\lambda_3\lambda_4}
 =\epsilon_{\mu}(p_1,\lambda_1)\epsilon_{\nu}(p_2,\lambda_2) \nn\\
 &\hspace*{1cm}\times\big( \mathcal{M}^{S,\mu\nu}_{\lambda_3\lambda_4}
         +\mathcal{M}^{P,\mu\nu}_{\lambda_3\lambda_4}
         +\mathcal{M}^{t,\mu\nu}_{\lambda_3\lambda_4}
         +\mathcal{M}^{u,\mu\nu}_{\lambda_3\lambda_4}\big),
\end{align}
where the photon wavefunctions are factorized. Using the straightforward
Feynman rules for Majorana fermions~\cite{Denner:1992vza}, the above
amplitudes are written, based on the effective Lagrangian in
Appendix~\ref{sec:lag}, as  
\begin{align}
 &i\mathcal{M}^{S,\mu\nu}_{\lambda_3\lambda_4}=
-\frac{im_{\nored}m^2_{\phi}}{F^2}\frac{1}{s-m_{\phi}^2}\,
 (p_1\cdot p_2\,g^{\mu\nu}-p_2^{\mu}p_1^{\nu}) \nn\\
 &\hspace*{1.5cm}\times\ubar(p_3,\lambda_3)v(p_4,\lambda_4),\\
 &i\mathcal{M}^{P,\mu\nu}_{\lambda_3\lambda_4}= 
-\frac{im_{\nored}m^2_{\phi}}{F^2}\frac{1}{s-m_\phi^2}\,
\epsilon^{\mu\nu\alpha\beta}\,{p_2}_{\alpha}{p_1}_{\beta} \nn\\
 &\hspace*{1.5cm}\times\ubar(p_3, \lambda_3)i\gamma^5v(p_4,\lambda_4),\\
 &i\mathcal{M}^{t,\mu\nu}_{\lambda_3\lambda_4}=
-\frac{im^2_{\nored}}{8F^2}\frac{1}{t-m^2_{\nored}} \nn\\
&\hspace*{0.6cm}\times\ubar(p_3,\lambda_3)[\gamma^{\mu},\slashed{p}_1] 
 (\slashed{p}_1-\slashed{p}_3-m_{\nored}) 
 [\slashed{p}_2,\gamma^{\nu}]v(p_4,\lambda_4), \\
 &i\mathcal{M}^{u,\mu\nu}_{\lambda_3\lambda_4}=
\frac{im^2_{\nored}}{8F^2}\frac{1}{u-m^2_{\nored}} \nn\\
&\hspace*{0.6cm}\times\ubar(p_3,\lambda_3)[\gamma^{\mu},\slashed{p}_2]
 (\slashed{p}_1-\slashed{p}_4+m_{\nored})
 [\slashed{p}_1,\gamma^{\nu}]v(p_4,\lambda_4),
\end{align}
where the common sgoldstino mass is taken as $m_{S,P}=m_{\phi}$. The
reduced helicity amplitudes $\hat{\mathcal{M}}$ are defined as  
\begin{align}
  \mathcal{M}_{\lambda_1\lambda_2,\lambda_3\lambda_4}
 =\frac{m_{\nored}s^{3/2}}{2F^2}\,
  \hat{\mathcal{M}}_{\lambda_1\lambda_2,\lambda_3\lambda_4},
\label{red_hel_amp}
\end{align}
and presented in Table~\ref{tb:helamp_aa}. The analytic expression for
the total cross section can be found in~\cite{Brignole:1996fn}, and 
our numerical results agree well with it.

\begin{table*}
\center
 \caption{The reduced helicity amplitudes
 $\hat{\mathcal{M}}_{\lambda_1\lambda_2,\lambda_3\lambda_4}$ defined
 in~\eqref{red_hel_amp} for
 $\gamma^{}_{\lam_1}\gamma^{}_{\lam_2}\to\gld^{}_{\lam_3}\gld^{}_{\lam_4}$.}
\begin{tabular}{c|cccccc}
\hline
 $\lambda_1\lambda_2\,\lambda_3\lambda_4$ & &
 $\hat{\mathcal{M}}^S$& $\hat{\mathcal{M}}^P$ &
 $\hat{\mathcal{M}}^t$& $\hat{\mathcal{M}}^u$ & \\\hline
 $\pm\pm\,\pm\pm$&
	 $\mp\Big[$&
	     $\dfrac{m^2_{\phi}}{s-m^2_{\phi}}$&
		 $-\dfrac{m^2_{\phi}}{s-m^2_{\phi}}$&&&$\Big]$ \\
 $\pm\pm\,\mp\mp$&
	$\pm\Big[$&
	     $\dfrac{m^2_{\phi}}{s-m^2_{\phi}}$&
		 $+\dfrac{m^2_{\phi}}{s-m^2_{\phi}}$& 
		    $ -\dfrac{m^2_{\nored}}{t-m^2_{\nored}}(1-\cos{\theta})$&
			$-\dfrac{m^2_{\nored}}{u-m^2_{\nored}}(1+\cos{\theta})$&$\Big]$ \\
 $\pm\mp\,\pm\mp$&
     &&&&$\dfrac{m_{\nored} \sqrt{s} }{
			 (u-m^2_{\nored})}\dfrac{1}{2}(1+\cos{\theta}) \sin{\theta}$& \\	
$\pm\mp\,\mp\pm$&
     &&&$-\dfrac{m_{\nored} \sqrt{s}}{
		     (t-m^2_{\nored})} \dfrac{1}{2}(1-\cos{\theta})\sin{\theta}$&& \\\hline
\end{tabular}
\label{tb:helamp_aa}
\end{table*}

\begin{figure}
\center
 \includegraphics[width=0.88\columnwidth]{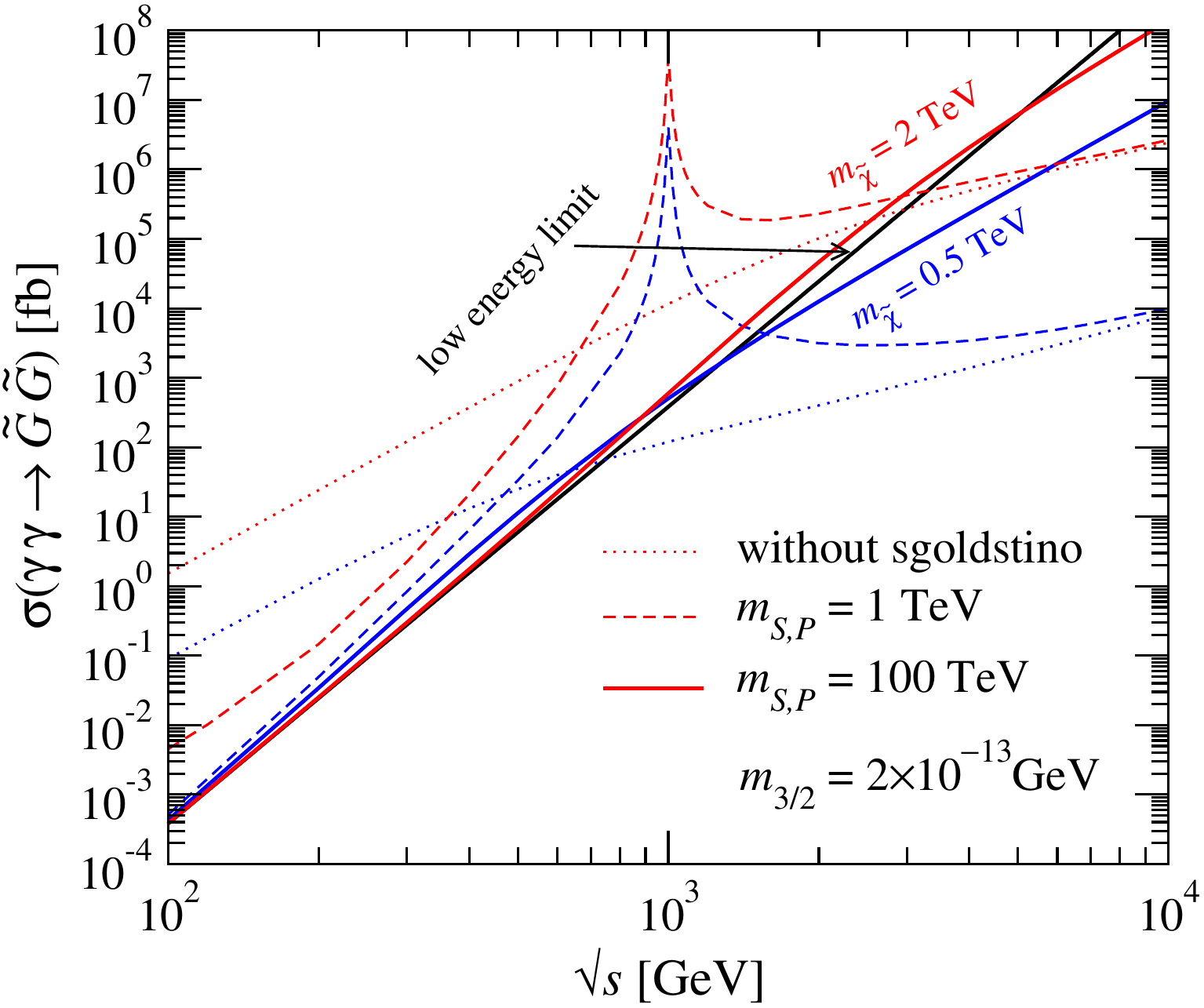}
 \caption{Total cross sections of $\gamma\gamma\to\gld\gld$ as a
 function of the collision energy for $m_{3/2}=2\times10^{-13}$~GeV. 
 The sgoldstino masses are taken to be 1~TeV (dashed) and 100~TeV
 (solid), while the neutralino mass is fixed at 0.5~TeV (blue) and 2~TeV
 (red). We also show the cross section in the low energy limit (black
 solid) as well as the contributions without the sgoldstino interactions
 (dotted).} 
\label{fig:sqrts_dep_aa_mSgold1TeV} 
\end{figure}

Figure~\ref{fig:sqrts_dep_aa_mSgold1TeV} shows the total cross sections
as a function of the CM energy $\sqrt{s}$ for $m_{\nored}=0.5$~TeV
(blue) and $m_{\nored}=2$~TeV (red) with
$m_{3/2}=2\times10^{-13}$~GeV. First, let us consider the heavy
sgoldstino case, $m_{\phi}=100$~TeV. In the low-energy limit,
$\sqrt{s}\ll m_{\phi,\nored}$, similar to the $e^+e^-$
collision~\eqref{s3}, the total cross section is given
by~\cite{Brignole:1996fn}  
\begin{align}
 \sigma=\frac{s^3}{640\pi F^4}, 
\label{xs_aa_low_energy}
\end{align}
shown by a black-solid line in Fig~\ref{fig:sqrts_dep_aa_mSgold1TeV}.  
Due to a cancellation between the sgoldstino and neutralino amplitudes
for $\lambda_1=\lambda_2=-\lambda_3=-\lambda_4$ as can be seen in
Table~\ref{tb:helamp_aa}, the dominant contribution is given by the
amplitudes for $\lambda_1=-\lambda_2$, which are proportional to $s^2$
in the low-energy limit. To emphasize the importance of the
interference, the contribution without the sgoldstino amplitudes is also
shown by a dotted line in Fig.~\ref{fig:sqrts_dep_aa_mSgold1TeV}. On the
other hand, in the case where the neutralino mass is smaller than the CM
energy, $m_{\nored}\ll\sqrt{s}\ll m_{\phi}$, the cross section is
dominated by the sgoldstino contributions and deviates from the one in
the low-energy limit. 
   
We now turn to the case where the sgoldstinos are relatively light,
$m_{\phi}=1$~TeV. In our SUSY QED model, the partial decay width of the
sgoldstinos are given by~\cite{Perazzi:2000id} 
\begin{align}
 \Gamma(S,P\to\gld\gld) &=\frac{m^5_{\phi}}{32 \pi F^2}, \\
 \Gamma(S,P\to\gamma\gamma) &=\frac{m^2_{\nored}m^3_{\phi}}{32\pi F^2}.
\end{align}
For $m_{\phi}=1$~TeV and $m_{3/2}=2\times 10^{-13}$~GeV
(i.e. $\sqrt{F}\approx 918$~GeV), the width for a gravitino pair is
14.0~GeV and for a photon pair is 3.5 (55.9)~GeV for $m_{\nored}=0.5$
(2)~TeV. For the $m_{\nored}=2$~TeV case, the finite width effect can be
seen as a deviation from the cross section~\eqref{xs_aa_low_energy} in
the low-energy region in Fig.~\ref{fig:sqrts_dep_aa_mSgold1TeV}. For
$\sqrt{s}\approx m_{\phi}$, one can clearly see the resonant peak. In
the high-energy limit, $\sqrt{s}\gg m_{\phi,\nored}$, the cross section
approaches the value obtained by neglecting the sgoldstino amplitudes,
since the $\lambda_1=-\lambda_2$ amplitudes become dominant; see
Table~\ref{tb:helamp_aa}.

Finally, we note that collider signatures of sgoldstinos have been
studied 
in~\cite{Dicus:1990su,Perazzi:2000id,Perazzi:2000ty,Gorbunov:2000th,Gorbunov:2000ht,Abreu:2000ij,Checchia:2001gd,Gorbunov:2001pd,Gorbunov:2002er,Demidov:2004qt,Petersson:2011in,Bellazzini:2012mh}, 
and our model file can be also applied for such sgoldstino phenomenology.

%RRRRRRRRRRRRRRRRRRRRRRRRRRRRRRRRRRRRRRRRRRRRRRRRRRRRRRRRRRRRR

\bibliography{library}
\bibliographystyle{JHEP} 

\end{document}